\providecommand\reserveinserts[1]{} % patch for Wiley class on arXiv

\documentclass[AMA,Times1COL]{WileyNJDv5} %STIX1COL,STIX2COL,STIXSMALL

\usepackage{graphicx} % Required for inserting images
\usepackage{amsmath}
\usepackage{amssymb}
\usepackage{bm}
\usepackage{fancyvrb}
\usepackage{appendix}
\usepackage{xcolor}
\usepackage{listings}
\lstdefinelanguage{Stan}{
  morekeywords={
    data, parameters, transformed, model, generated, quantities,
    if, else, for, while, break, continue, return, 
    real, int, vector, simplex, matrix, cov_matrix, cholesky_factor_corr, 
    cholesky_factor_cov, ordered, positive_ordered, unit_vector, 
    row_vector, array, lower, upper, target
  },
  sensitive=true,
  morecomment=[l]{//},       % line comment
  morecomment=[s]{/*}{*/},   % block comment
  morestring=[b]"            % strings
}

\definecolor{ForestGreen}{rgb}{0.0,0.4,0.0}

\lstdefinelanguage{R}{
  keywords={if, else, repeat, while, function, for, in, next, break, summary},
  sensitive=true,
  morecomment=[l]{\#},     % line comments
  morestring=[b]",         % strings
  morestring=[d]'          % strings with single quotes
  language=R,
  basicstyle=\fontsize{9}{11}\selectfont\ttfamily,
  keywordstyle=\color{black},
  commentstyle=\color{gray}\bfseries,
  stringstyle=\color{ForestGreen}\bfseries,
  numberstyle=\tiny\color{gray}\bfseries,
  numbers=left,
  stepnumber=1,
  numbersep=5pt,
  tabsize=2,
  showstringspaces=false,
  breaklines=true,
  frame=single,
  captionpos=b
}

\newcommand{\stan}{\texttt{Stan} }

\newcommand{\cd}[1]{\texttt{#1}}
\newcommand{\expit}{\text{expit}}
\newcommand{\Eta}{ \text{H} }

\articletype{}%

\startpage{1}

\begin{document}

\title{Stress-Testing Assumptions: A Guide to Bayesian Sensitivity Analyses in Causal Inference}

\author[1]{Arman Oganisian}

\authormark{Oganisian}
\titlemark{Bayesian Sensitivity Analyses in Causal Inference}

\address[1]{\orgdiv{Department of Biostatistics}, \orgname{Brown University}, \orgaddress{\state{Rhode Island}, \country{United States}}}

\corres{Corresponding author Arman Oganisian. \email{arman\_oganisian@brown.edu}.}

\presentaddress{121 South Main Street, Providence, RI, USA 02903}

%\fundingInfo{Text}
%\JELinfo{ejlje}

\abstract[Abstract]{While observational data are routinely used to estimate causal effects of biomedical treatments, doing so requires special methods to adjust for observed confounding. These methods invariably rely on untestable statistical and causal identification assumptions. When these assumptions do not hold, sensitivity analysis methods can be used to characterize how different violations may change our inferences. The Bayesian approach to sensitivity analyses in causal inference has unique advantages as it allows users to encode subjective beliefs about the direction and magnitude of assumption violations via prior distributions and make inferences using the updated posterior. However, uptake of these methods remains low since implementation requires substantial methodological knowledge. Moreover, while implementation with publicly available software is possible, it is not straight-forward. At the same time, there are few papers that provide practical guidance on these fronts. In this paper, we walk through four examples of Bayesian sensitivity analyses: 1) exposure misclassification, 2) unmeasured confounding, and missing not-at-random outcomes with 3) parametric and 4) nonparametric Bayesian models. We show how all of these can be done using a unified Bayesian ``missing-data'' approach. We also cover implementation using \stan, a publicly available open-source software for fitting Bayesian models. To the best of our knowledge, this is the first paper that presents a unified approach with code, examples, and methodology in a three-pronged illustration of sensitivity analyses in Bayesian causal inference. Our goal is for the reader to walk away with implementation-level knowledge.
}


\jnlcitation{\cname{%
\author{Oganisian A.},
\ctitle{On simplifying ‘incremental remap’-based transport schemes.} \cjournal{\it J Comput Phys.} \cvol{2021;00(00):1--18}.}}

\maketitle

\renewcommand\thefootnote{}

\renewcommand\thefootnote{\fnsymbol{footnote}}
\setcounter{footnote}{1}

\section{Introduction}

In the absence of well-designed and controlled trials with randomized treatments, researchers often turn to observational data to estimate the effects of treatment decisions. A cost of using observational data is that sophisticated causal inference methods must be used to adjust for observed confounding. This has led to a large literature on causal inference methods over the past several decades. All such methods, however, rely on untestable assumptions - either causal assumptions (i.e. assumptions about the treatment mechanism) or the statistical assumptions about the data observation mechanism. For example, estimation of causal effects of point treatments requires that the treatment assignment mechanism is unconfounded. In settings where we observe data incompletely (e.g. with error-prone measurements, missing values, etc.), we require more untestable assumptions about the mechanism by which the missing data was peeled off from the complete data to leave us with the observed data at hand. This is sometimes known as the ``missing data mechanism'' \cite{molenberghs2014}. 

Untestable assumptions are unavoidable in practical settings. Indeed, ``Nothing is wrong with making assumptions...It is the scientific quality of those assumptions, not their existence, that is critical.'' \cite{Rubin2005}. The goal of sensitivity analyses is precisely to help assess the scientific quality and impact of these assumptions. Recently, Cinelli et al. (2025)\cite{cinelli2025} enumerated twelve open challenges in causal inference with the sixth being ``Sensitivity Analysis and Robustness.'' They first note that ``such methods have not yet been widely adopted in the empirical literature.'' Second, they remark that while much work has been done assessing sensitivity to unmeasured confounding, ``other types of biases may also pose substantial threats to valid causal inferences. For examples, the problems of sample selection, attrition, missing data, measurement error'' - alluding to the aforementioned missing-data mechanism mentioned before. This paper seeks to address these issue by providing a didactic walkthrough of several examples of sensitivity analyses alongside publicly available code\footnote{All code referenced in this manuscript can be found here: \url{https://github.com/stablemarkets/StressTest}} within the Bayesian framework. In order to provide a unifying framework, we provide examples that go beyond assessing sensitivity to ignorability violations and discuss violations of other assumptions as well. 

We work within a Bayesian framework for causal inference, which is becoming increasingly popular. Interested readers can now find many published reviews and textbooks \cite{Oganisian2021a,li2023,linero2023,daniels2023bayesian} on the topic at various levels ranging from introductory to advanced. This body of work exploits four benefits to the Bayesian approach to causal inference: 1) the availability of an array of flexible Bayesian nonparametric methods that reduce dependence on model misspecification, 2) priors as a form of regularization/shrinkage, and 3) full posterior uncertainty quantification even with nonparametric models and intricate prior regularization structures.

A fourth benefit of the Bayesian approach - which makes it uniquely useful for sensitivity analysis - is a lack of fundamental distinction between ``parameters'' and ``data'' in a given model. In Bayesian inference, both quantities are random variables and are distinguished only in that some are unknown or while others are known during the analysis. Parameters are random variables which we do not know (i.e. do not have realizations of) whereas the observed data are random variables we know (i.e. do have a realization of at hand). Bayesian inference in all settings - whether for causal effects, missing data problems, longitudinal model with latent random effects, etc. - is done with the same procedure: we make inferences about unknown quantities conditional on known quantities. That is, we find the posterior distribution. Since untestable assumptions are beliefs about quantities we do not know, sensitivity analysis can be done by treating these quantities simply as additional unknowns about which we make posterior inferences.

This makes the Bayesian paradigm a unifying framework for assessing the sensitivity of our causal effect estimates to untestable assumptions. This has been well-known in the Bayesian methodological literature but has not seen wide adoption in practice. For example,  Bayesian sensitivity analysis methods have been explored for unmeasured\cite{McCandless2007} or mismeasured\cite{Gustafson2010} confounders as well as unmeasured confounders in more complex mediation settings \cite{McCandless2019}. Other work has developed sensitivity analysis for unmeasured confounding in settings with recurrent, truncated outcomes \cite{Sisti2024} and with time-varying treatments \cite{zou2025}. Others have outlined such Bayesian sensitivity analyses for unmeasured confounding within the context of nonparametric models\cite{Roy2016}. Outside of an explicit causal context, Bayesian sensitivity analyses for related missing data problems have also been developed \cite{daniels2008missing,10.1214/17-STS630,molenberghs2014}

Lack of wide-scale adoption is primarily due to two factors. First, despite advances in the methodological literature, there is a lack of accessible material available for practitioners. Second methodological work is often implemented using tailored Bayesian computation algorithms for the purposes of running benchmark simulation studies, rather than using computational tools used in practice. It is not commonly known that such methods can, in fact, be implemented with off-the-shelf, publicly available software. While probabilistic programming languages (PPLs) such as \stan \cite{stan2017} exist for specifying Bayesian models and have broad use, doing causal sensitivity analyses in \stan is not straightforward. There remains a gap of accessible resources that combine methodological knowledge with concrete implementation guidance. 

In order to address these gaps, this paper presents four examples of sensitivity analyses in Bayesian causal inference. The first is a sensitivity analysis for imperfectly ascertained treatment assignments (a.k.a ``exposure misclassification''), the second is a sensitivity analysis for unmeasured confounding, and the third is sensitivity analysis for missing not-at-random (MNAR) outcomes. In these three examples we use routine, parametric Bayesian models to keep focus on the implementation and sensitivity analysis concepts. However, since a major appeal of Bayesian causal inference is the availability nonparametric models, we conclude with a fourth example using flexible stick-breaking mixture models. For each example, we include a demonstration running \stan on synthetic data sets which we make publicly available in our companion GitHub repository. This is the first paper to our knowledge that presents code, examples, and methodology in a three-pronged approach to build intuition on sensitivity analyses in Bayesian causal inference. Our goal is for the reader to walk away with implementation-level knowledge of Bayesian sensitivity analysis and sufficient intuition to do such analyses in their own causal inference work. 

We begin in Section 2 by reviewing Bayesian causal inference in the ideal case where all assumptions hold. We also provide a review of Bayesian causal computation in \stan and the basics of specifying custom likelihood and priors within \stan, which is essential in coding sensitivity analyses. Section 3 beings to relax the base-case developed in Section 2 - with each of the four sub-sections devoted to the first three examples mentioned above. Section 4 discusses an example with nonparametric models.

\section{Review of Bayesian Causal Inference when Assumptions Hold}
\label{sc:bayesian_causal_standard}

Throughout this paper, we consider a point-treatment setting with a single outcome. In this setting, we ideally observe data $D^O=\{y_i, a_i, l_i\}_{i=1}^n$ consisting of an outcome $Y_i$, a binary treatment indicator $A_i$, and a length-$P$ covariate vector $L_i$ for $n$ subjects indexed by $i=1,2,\dots, n$. Throughout, we use capital letters for random variables and lowercase for realization in both Roman script (e.g. $Y$ versus $y$) and Greek script (e.g. $\Theta$ and $\Delta$ versus $\theta$ and $\delta$). We omit subscript $i$ when discussing an arbitrary subject for compactness.

For each subject we define one potential outcome \cite{Neyman1923} for each value of the treatment of interest. In the binary setting, we have $Y^a$ for $a\in\{0,1\}$. Population-level causal effects are defined as comparisons of expected potential outcomes within a common population. For instance, we will focus on the average treatment effect $ATE$, $\Psi = E[Y^1 - Y^0]$. This is the difference between the expected outcome in the target population had we intervened to treat everyone with treatment $A=1$ versus $A=0$. Hereafter, for simplicity we may sometimes refer to $A=0$ as ``untreated'' or ``no treatment'' though treatment option $A=0$ may be active in general. In observational studies, $\Psi$ is not in general equal to the associational estimand $E[Y\mid A=1] - E[Y\mid A=0]$, which contrasts expected \textit{observed} (not potential) outcomes between those who \textit{happened to be} treated with those who \textit{happened to be} untreated. This is because in observational studies, covariates $L$ may affect both the outcome and selection into treatment such that, say, those who happened to get treatment $A=1$ are not representative of the target population. Thus we call such covariates ``confounders'' because differences in observed outcomes between treatment groups cannot be teased out from differences in $L$ between these groups.

All causal inference requires assumptions because while we wish to make inferences about $\Psi$, neither potential outcome is present in the observed data, $D^O$. These causal identification assumptions link the distribution of the potential outcomes to the distribution of the observed data.

The stable unit treatment value (SUTVA) assumption asserts that $Y = AY^1 + (1-A)Y^0$, which explicitly links the observed outcome with the \textit{factual} potential outcome. It implies that for a subject with observed treatment $a_i$, $Y_i = Y^{a_i}$. Even under SUTVA, however, half of the potential outcomes - the \textit{counterfactuals}, $Y_i^{1-a_i}$ - are still unobserved and, so, more assumptions are needed about the treatment mechanism. Once such assumption is that the treatment mechanism is ``unconfounded'' - that is, $A \perp Y^1, Y^0 \mid L=l$ or equivalently, $P(A=1 \mid L=l, Y^1, Y^0) = P(A = 1\mid L=l)$. This is sometimes also referred to as the ``exchangeability'' assumption as it asserts that treated and untreated patients who have the same $L$ are exchangeable with respect to the distribution of their potential outcomes. It is also sometimes referred to as ``no unmeasured confounding'' because it asserts that the measured covariates $L$ are sufficient for this conditional independence to hold. Additionally, we must assume that this treatment assignment mechanism is ``probabilistic'' - i.e. $0<P(A=1\mid L=l)<1$ for each value of the covariate $l$ where $f_L(l)>0$. This assumption is also variously called the ``overlap'' assumption or ``positivity'' assumption. A treatment assignment that is both ``unconfounded'' and ``probabilistic'' is said to be ``strongly ignorable''\cite{Rubin2007b}. For strongly ignorable treatment assignments, the ATE can be identified as the difference in conditional (on $L$) expected outcomes between the two treatment groups, averaged over the population distribution of the covariates,

\begin{equation}\label{eq:gcomp}
    \Psi(\eta,\theta) = \int \Big( E[Y\mid A=1, L=l;\eta] - E[Y\mid A=0, L=l;\eta]\Big) f_L(l;\theta) dl
\end{equation}

Because the right-hand side of the equality involves only functionals of the distribution of $D^O$, we say the ATE is ``identified.'' This formula is known as standardization and is a special case of the g-formula \cite{ROBINS1986}, which generalizes standardization from point-treatment to time-varying treatment settings. Above, $\eta$ and $\theta$ are parameters governing the distribution of the outcome and covariates, respectively. Models in which these parameters have have low, fixed dimensionality are often called parametric models. In the Bayesian literature, models in which parameters are high-dimensional - perhaps with dimensionality growing with $n$ - are often called ``nonparametric.'' We write the ATE as $\Psi=\Psi(\eta, \theta)$ to emphasize that it is a function of these unknown parameters. One method of doing frequentist inference for the ATE is to solve unbiased estimating equations to obtain $(\hat \eta,  \hat \theta)$, then plugging these into \eqref{eq:gcomp} to get $\hat \Psi = \Psi(\hat \eta, \hat \theta)$. This involves computing the integrals in \eqref{eq:gcomp}, which in practice may be done with  Monte Carlo simulation. Inferences can be made by repeating the process across bootstrap resamples to obtain, say, a 95\% confidence interval for $\Psi$.

In contrast, since $\Psi(\eta,\theta)$ is a functional of unknown parameters, Bayesian inferences are made using the posterior distribution $f(\eta, \theta\mid D^O)$, which induces a posterior over $\Psi$. This requires 1) models for the outcome and covariate distribution and 2) the use of Markov Chain Monte Carlo (MCMC) methods for sampling from the posterior over these parameters. The latter is necessary because even with low-dimensional/parametric models, the posterior $f(\eta, \theta\mid D^O)$ will not be a known distribution with a closed form.

We will now discuss these in the arbitrary case and, in the subsequent section, discuss a concrete implementation example. Let $f_{Y|A,L}(y \mid a, l; \eta )$ and  $f_L(l; \theta)$ denote the models for the conditional outcome  and the marginal covariate probability density/mass functions (pdfs/pmfs), governed by parameters $\eta$ and $\theta$ respectively - which for simplicity we assume are continuous and real-valued. We let $f_\Eta(\eta)$ and $f_\Theta(\theta)$ denote the prior pdfs on these parameters, which must also be specified. Once these models and priors are specified, Bayes' rule dictates that the posterior, up to a proportionality constant, is 

\begin{equation} \label{eq:post_density}
    f(\eta, \theta\mid D^O) \propto C\cdot f_\Eta(\eta)f_\Theta(\theta)\prod_{i=1}^n f_{Y|A,L}(y_i \mid a_i, l_i; \eta ) f_L(l_i; \theta)
\end{equation}

Note that we do not know the posterior density to an equality - we only know it is proportional to $\tilde f(\eta, \theta \mid D^O) = f_\Eta(\eta)f_\Theta(\theta)\prod_{i=1}^n f_{Y|A,L}(y_i \mid a_i, l_i; \eta ) f_L(l_i; \theta)$. The tilde in $\tilde f$ distinguishes the \textit{unnormalized} posterior density from the posterior density $f(\eta, \theta\mid D^O)$. Another important note is that, in the above, $C= \int \prod_{i=1}^n f_{A|L}(a_i\mid l_i; \gamma)f_\Gamma(\gamma) d\gamma$ is the treatment model contributions averaged over its parameters' prior distribution. As is commonly done, we assume here that the outcome, covariate, and treatment models are governed by distinct \textit{a priori} independent parameters and thus $C$ can be ignored (i.e. treated as $C=1$) while maintaining proportionality in \eqref{eq:post_density}. We shall see that in certain sensitivity analysis examples, however, the unknowns of interest will depend on the treatment model and so it cannot be ignored.

In practical settings, even though $ f(\eta, \theta\mid D^O) $ is unavailable, $\tilde f(\eta, \theta \mid D^O)$ is available: it is simply the product (over the observed data) of the prior and model pdfs/pmfs we specify. Fortunately, MCMC only requires $\log \tilde f$ to generate a representative set of draws from the posterior $f(\eta, \theta \mid D^O)$. Suppose for now we can use MCMC to obtain a set of $M$ posterior draws $\{\eta^{(m)}, \theta^{(m)} \}_{m=1}^M$. Then, the $m^{th}$ posterior draw of $\Psi$ is given by plugging in the $m^{th}$ draw of these parameters into the standardization formula

$$ \Psi^{(m)} = \int \Big( E[Y\mid A=1, L=l;\eta^{(m)}] - E[Y\mid A=0, L=l;\eta^{(m)}]\Big) f_L(l;\theta^{(m)}) dl$$

Doing this for all $M$ draws and taking the average yields an approximation to the posterior mean of $\Psi$. Taking, say, the 2.5th and 97.5th percentiles of these $M$ draws yields a 95\% credible interval for $\Psi$. Since all the usual general guidance about Bayesian model-checking, diagnostics, and MCMC convergence checks apply here, we refer users to existing work on these issues \cite{gelman2013bayesian, andrieu2003}. We do this in order to keep focus on the required causal and sensitivity analysis concepts.

\subsection{Bayesian Causal Computation Using \stan} \label{sc:stan_review}

When making Bayesian inferences about $\Psi$, it is infeasible to code MCMC samplers from scratch on a problem-by-problem basis to obtain draws $\{\eta^{(m)}, \theta^{(m)} \}_{m=1}^M$. To address this gap, probabilistic programming languages (PPLs) can be used. PPLs provide a syntax with which analysts may specify a \textit{log} unnormalized posterior density $\log \tilde f(\eta, \theta\mid D^O)$, comprised of a log likelihood function and the log prior density function. The PPL software will then run MCMC methods under the hood and return a specified number of draws from the corresponding posterior.

Though there are many PPLs (e.g., Greta, PyMC3, Nimble, JAGS, BUGS, etc.), we will focus implementation in \stan, which is perhaps the most common PPL used in Bayesian causal inference. One advantage of \stan is that popular Bayesian inference textbooks\cite{gelman2013bayesian} are written around Stan and there is a large and active online community of users. Another advantage is that, while there are many MCMC algorithms, \stan runs a variant of Hamiltonian Monte Carlo (HMC) under the hood \cite{neal2011mcmc,betancourt2017}. HMC is a gradient-based, adaptive MCMC method that can generate draws efficiently even in models with high-dimensional parameters exhibiting complex posterior dependence structures.

\begin{figure}[h!]
\begin{minipage}[t]{.48\linewidth} \scriptsize
\begin{lstlisting}[title={},
                   caption={Excerpt from \cd{gcomp\_complete.R} file. },
                   language=R, captionpos=t]
library(rstan)
...
### --- Create Stan Data --- ###
## n is scalar integer sample size
## y, a, l are length-n vectors 
## of outcome, treatment, and covariate data
stan_data = list(n=n, 
                 y=y, 
                 a=a, 
                 l=l)
 
### --- Load the Stan Model --- ###
mod = stan_model("gcomp_complete.stan")

## Obtain a single chain
## with 1k posterior draws (after 1k warm-up draws)
res = sampling(mod, 
               seed=31, 
               data=stan_data,
               chains = 1, 
               iter = 2000, 
               warmup = 1000)

## extract posterior summary of ATE
draws = summary(res, pars=c('ATE') )$summary
\end{lstlisting}
\end{minipage}
\hfill
\begin{minipage}[t]{.48\linewidth} \scriptsize
\begin{lstlisting}[title={},captionpos=t,
                   caption={Excerpt from \cd{gcomp\_complete.stan} file.}]
data {
  // declare known quantities (data)
  int<lower=0> n;
  int y[n]; int a[n]; int l[n];
}

parameters {
  // declare unknown quantities (parameters)
  real<lower=0, upper=1> theta;
  real eta[3];
}


model{
  theta ~ beta(1,1);
  eta ~ normal(0,3);

  for(i in 1:n){
    y[i] ~ bernoulli(inv_logit( eta[1] + eta[2]*l[i] + eta[3]*a[i]));
    l[i] ~ bernoulli(theta);
  }
}

generated quantities {
  // standardization
  real ATE;
  
  ATE = theta*( inv_logit(eta[1]+eta[2]+eta[3]) 
                - inv_logit(eta[1]+eta[2])) +
    (1-theta)*( inv_logit(eta[1]+eta[3]) 
                - inv_logit(eta[1]) ); 
}
\end{lstlisting}
\end{minipage}
\caption{ The \cd{.R} (Left) and \cd{.stan} (Right) files implementing the causal computation procedure described in Section \ref{sc:bayesian_causal_standard} with a binary outcome and $P=1$ binary covariate. Ellipsis indicates lines skipped for compactness; the full code can by found in the companion GitHub repository. We specified $f_{Y|A,L}(y\mid a, l; \eta)$ as a Bernoulli pmf with conditional probability $ P_{Y|A,L}(Y=1\mid A=a, L=l; \eta) = \expit ( \eta_0  +\eta_1l + \eta_2a)$.
For the covariate model, we specify $f_L(l;\theta)$ as a Bernoulli pmf with probability $0<\theta<1$.}
\label{fig:stan_gcomp}
\end{figure}

\stan is a distinct programming language that interfaces with other programming languages through packages such as \cd{rstan} \cd R or \cd{pystan} in \cd{Python}. It can even be called directly from the terminal. We will focus on interfacing with \cd{rstan}. A common workflow, which we also adopt for all example code accompanying this paper, is to write two separate files in the same working directory:
\begin{enumerate}
	\item \cd{.stan} file: contains  \texttt{Stan} syntax that specifies likelihood and priors - i.e. the components of the log unnormalized posterior density, $\log \tilde f$, of a Bayesian model.
	\item \cd{.R} file: contains \cd{R} code that manipulates data and, using \cd{rstan}, calls the Bayesian model specified in the \cd{.stan} file and passes the data to it in order to it.
\end{enumerate}
The posterior draws are then returned in the \cd{R} session and can be manipulated just like all other \cd R objects. \cd{rstan} also provides convenience functions for summarizing and visualizing posterior draws as well as assessing MCMC diagnostics.

The \cd{.stan} file is organized in terms of the following ``blocks'':
\begin{enumerate}
	\item \cd{data} block: declares and specifies the observed data structures, e.g. covariates, outcomes, sample size, etc. 
	\item \cd{transformed data} block (optional): declares and specifies desired transformations of the objects declared in data block.
	\item \cd{parameters} block: declares and specifies the structure of the parameters in the model.
	\item \cd{transformed parameters} (optional): declares and specifies structure of desired transformations of the declared parameters.
	\item \cd{model} block: specifies likelihood and prior distributions.
	\item \cd{generated quantities} block (optional): specifies transformations of the parameters for which you would also like to return posterior draws. If not specified, default is to return draws of all quantities declared in the ``parameters'' and ``transformed parameters'' blocks.
\end{enumerate}

Figure \ref{fig:stan_gcomp} shows the \cd{.R} and \cd{.stan} file implementing the causal computation procedure described in Section \ref{sc:bayesian_causal_standard} with a binary outcome and $P=1$ binary covariate. We specified a logistic model for $f_{Y|A,L}(y\mid a, l; \eta)$. That is, $f_{Y|A,L}(y\mid a, l; \eta)$ is a Bernoulli pmf with conditional probability 
$$ P_{Y|A,L}(Y=1\mid A=a, L=l; \eta) = \expit ( \eta_0  +\eta_1l + \eta_2a)$$
We specify the covariate model, $f_L(l;\theta)$, to be a Bernoulli pmf with probability $0<\theta<1$. We stress that the code we present here is optimizing for didactic clarity, even at the cost of computational efficiency. We try to match the code with the mathematical notation as closely as possible to build intuition, but more efficient implementations are possible.

The \cd{data} block of the \stan file declares four variables: the sample size \cd n, as well as length \cd n arrays for the outcome, treatment and covariate. Note that unlike \cd R which is a weak, dynamically typed language, \stan is a strong, statically typed language. That is, each variable has a type, size, and range declared that cannot be changed later in the program (static typing). If operations are performed between variables of inappropriate types, \stan will throw an error and stop execution rather than coerce the types to make the operation work (strong typing). The type, size, and constraints for the variables are specified to obey probabilistic constraints. For example, the \cd{parameters} block of the \cd{.stan} file in Figure \ref{fig:stan_gcomp} declares a length 3 vector \cd{eta} which corresponds to $\eta=(\eta_0, \eta_1, \eta_2)$ in our logistic outcome regression (note \stan uses one-indexing, not zero-indexing). Thus, they are allowed to be unconstrained real numbers. It also declares a scalar parameter \cd{theta} that is a real number between 0 and 1: \cd{real<lower=0, upper=1>}. This constraint is appropriate since $\theta$ is the probability parameter in a Bernoulli model for the binary confounder.

The \cd{model} block of the \stan file specifies contributions to the log unnormalized posterior density of our Bayesian model. For this particular choice of models and priors, this is found by substituting our model pmfs/pdfs into the general form given on the right-hand side of Equation \ref{eq:post_density} to get:
\begin{equation*}
    \begin{split}
        \log \tilde f(\eta, \theta\mid D^O) & = \log Beta(\theta;1,1) + \log N_3(0_3;I) \\
        & +\sum_{i=1}^n \log Ber(y_i ; \expit(\eta_0+\eta_1l_i+\eta_2a_i)) + \log Ber(l_i;\theta)\\
    \end{split}
\end{equation*}
Above, and throughout the paper, the notation $``distribution(c;d)"$ denotes the density/mass function of the ``distribution'' governed by parameters $d$ and evaluated at $c$. For example, when we write \Verb+theta ~ beta(1, 1);+ in the \cd{model} block, it is actually instructing \stan to add $\log Beta(\theta;1,1)$ to the log unnormalized posterior density. \stan has many built-in distributions which can be found in its online manual. \\

Since Equation \ref{eq:gcomp} identifies the ATE, $\Psi$, as just a transformation of the outcome and covariate parameters, we compute a draw of the ATE in the \cd{generated quantities} block in Figure \ref{fig:stan_gcomp}. For this particular choice of model, Equation \ref{eq:gcomp} reduces to

$$ \Psi = \sum_{l=0}^1 \Big( \expit( \eta_0+\eta_1l + \eta_2) - \expit( \eta_0+\eta_1l) \Big)\theta^l(1-\theta)^{1-l} $$

Which is exactly what is computed in this block. Note that, in this case, no Monte Carlo integration is needed to evaluate the integral over $f_L(l;\theta)$. Due to the one-dimensional, binary nature of $L$ the integral is replaced with a sum of two terms each of which can be computed exactly.

We also note that \stan is a compiled language that interfaces through \cd R via the package \cd{rstan}. The \cd{stan\_model} function in the \cd{.R} excerpt in Figure \ref{fig:stan_gcomp} compiles the \stan script into \cd{C++} using the \cd{stanc} compiler under the hood. The \cd{C++} code is then compiled to binary and executed when running the \cd{sampling} function in the \cd{.R} file. The \cd{sampling} call in the \cd{.R} files instructs \stan to return 1,000 draws of all unknowns - quantities in the \cd{parameters} block and \cd{generated quantities} block - and store them in the object \cd{res}. This can then be summarized by, say, means and percentiles to form posterior point and interval estimates respectively. That is, \cd{res} contains the draws $\{\Psi^{(m)}\}_{m=1}^M$ for $M=1000$. The \cd{summary} function returns the mean, 2.5th, and 97.5th percentiles of these draws among other summary statistics. 

Part of the conclusion here is that Bayesian models are intuitive to specify in \stan as the log unnormalized posterior density contributions are coded analogously to how one would write them in mathematical notation.

\subsubsection{Custom Unnormalized Posterior Density Incrementation} \label{sc:custom}

In the previous sub-section we discussed that, in the \cd{model} block, specification of \Verb+theta ~ beta(1, 1)+ in the \cd{model} block, is instructing \stan to add $\log Beta(\theta;1,1)$ to the log unnormalized posterior density, $\log \tilde f(\eta, \theta \mid D^O)$. This syntax works well when contributions are additive on a log-scale and take the form of standard distributions for which \stan has pre-built functions, such as \cd{beta} and \cd{bernoulli} in Figure \ref{fig:stan_gcomp}.  In sensitivity analyses for causal inference, we will often need to specify non-standard, custom contributions to $\log \tilde f(\eta, \theta \mid D^O)$.

\stan can accommodate this via the keyword \cd{target}, which is its internal name for $\log \tilde f(\eta, \theta \mid D^O)$. Contributions can be added to this quantity via the compound assignment operator, \cd{+=}.\footnote{Note:\cd{x+=y} is the same as \cd{x = x + y}.}. For example, the \cd{model} block in the \cd{.stan} file of Figure \ref{fig:stan_gcomp} is equivalent to the following:

\begin{lstlisting}[title={},captionpos=t,
                   caption={},numbers=none,
                   label=labb]
model{
  target += beta_lpdf(theta | 1,1);
  target += normal_lpdf(eta | 0,3);

  for(i in 1:n){
   target += bernoulli_lpmf( y[i] | inv_logit( inv_logit( eta[1] + eta[2]*l[i] + eta[3]*a[i]) ) );
   target +=  bernoulli_lpmf( l[i] | theta );
  }
}
\end{lstlisting}

The functions with suffixes of the form \cd{\_lpdf} and \cd{\_lpmf} are optimized, numerically stable implementations of the corresponding log-density and log-mass functions, respectively. That is, in the code excerpt above, the following three statements specify the same contribution of $l_i$'s likelihood evaluation to the log unnormalized posterior:
\begin{itemize}
    \item \Verb& l[i] ~ bernoulli(theta);&, 
    \item \Verb& target +=  bernoulli_lpmf( l[i] | theta );&
    \item \Verb& target += l[i]*log(theta) + (1-l[i])*log(1-theta);&
\end{itemize}

This allows us to specify custom increments to the log unnormalized posterior density. However, whenever possible, we should increment \cd{target} using built-in functions like \cd{bernoulli\_lpmf} as in the second bullet rather than hand-coding log-pmfs and log-pdfs as in the third bullet. The built-in functions are more numerical stability and this can materially improve speed and performance of the sampler.

\section{Sensitivity Analyses in Causal Inference}

Bayesian inference and computation for $\Psi$ as described in Section \ref{sc:bayesian_causal_standard} crucially relies on both causal identification assumptions \textit{and} statistical assumptions. An example of the former is the unconfoundedness assumption that $A \perp Y^1, Y^0  \mid L$. However, in many settings $A \not \perp Y^1, Y^0  \mid L$ because there is a pre-treatment covariate, $U$, that impacts both treatment assignment and the outcome and yet is missing for all subjects. Such a $U$ is sometimes called an ``unmeasured confounder'' in the epidemiological literature.  An example of the latter is that it takes for granted that the treatment assignment indicator $A$ is ascertained without error. In the epidemiological literature this is sometimes referred to as ``no exposure misclassification.'' In many settings, however, we only observe an error-prone ascertainment, $\tilde A$, of a correct exposure status $A$. As another example, it assumes that the outcome $Y$ is measured for all subjects. In many cases, however, the outcome is available for some subjects and missing for others. Patients with observed outcome information may not be representative of the target population. 

Thus, in general the results from the approach in Section \ref{sc:bayesian_causal_standard} are sensitive to violations of these assumptions. Sensitivity analysis methods allow us to characterize the degree of this sensitivity and describe how results change across different beliefs about the direction and magnitude of assumption violations. In each of these three examples, violations of assumptions are due to the existence of missing data, $D^M$, that are not in $D^O$ - whether it is the correct treatment status $A$, the missing covariate, $U$, or the outcome values $Y$ for subjects who had it missing. As stated earlier, there is no distinction in Bayesian inference between parameters and data. Rather, both are random variables and the key distinction is between quantities that are observed (the data realization) versus unobserved (parameters and missing data). Thus, in the presence of missingness, Bayesian inference for $\Psi$ follows from simply viewing the missing quantities, $D^M$, and parameters that govern its distribution as additional unknowns and sampling from the larger \textit{joint} posterior of parameters \textit{and} missing quantities using \stan. That is, rather than sampling from the posterior $f(\eta, \theta\mid D^O)$ as in Section \ref{sc:bayesian_causal_standard}, in cases of violations we sample from the joint posterior  

$$ f(D^M, \xi, \eta, \gamma, \theta \mid D^O)$$

As before, $\eta$, $\theta$, and $\gamma$ are parameters that partially govern the outcome, covariate, and treatment model respectively. The quantity $\xi$ is the non-identifiable parameter vector that may govern some or all over these distributions, depending on the specifics of the problem. These additional parameters are known as ``sensitivity parameters'' and encode departures from standard assumptions about the missingness; these include for example, departures from ``no exposure misclassification'' or ``no unmeasured confounding'' or outcomes that are ``missingness-at-random''. Specified values of these sensitivity parameters - which we will call ``null values'' - encode that the assumptions are satisfied. We call values that encode departures from the assumptions as ``non-null values.'' Roughly, the steps for doing Bayesian sensitivity in causal inference can be summarized as follows:
\begin{enumerate}
    \item Identify $\Psi$ in terms of functionals of the complete-data distribution - i.e. functionals of the joint distribution over the missingness mechanism, the missing data, and the observed data. 
    \item Model the complete-data distribution components (i.e. the missingness mechanism, the missing data, and the observed data) using a mix of identifiable parameters and specified values of sensitivity parameters. 
    \item Use \stan to obtain draws from the joint distribution over all unknowns.
    \item Plug in the draws of the requisite parameters into the identified expression from Step 1 to obtain a draw of $\Psi$ as described in Section \ref{sc:bayesian_causal_standard}.
    \item Repeat Steps 2-4 under a range of specified null and non-null sensitivity parameter values to assess how posterior inference for $\Psi$ changes.
\end{enumerate}

The precise factorization of the complete-data posteriors, specification of sensitivity parameters, and modeling choices are application specific and also specific to the particular assumption violation we are exploring. Each of the next three sub-sections will illustrate a Bayesian sensitivity analysis for a different assumption violation with the goal of building intuition for modeling and implementation in \stan.

\subsection{Example 1: Exposure Misclassification} \label{sc:ex_misclass}
In the case of exposure misclassification, the true exposure status for each individual is missing $D^M =\{A_i\}_{i=1}^n$ and the observed data consists of $D^O = \{\delta_i=1, y_i, \tilde a_i, l_i \}$. Here, $\tilde A$ is the error-prone ascertainment of $A$ and $\delta_i$ is a missing-data indicator, the conditional distribution of which is often called the ``missingness mechanism''\cite{daniels2008missing,molenberghs2014}. In this example, $\delta_i$ is equal to one for all $i$ since we assume exposure is misclassified for all subjects and so we can ignore the missingness mechanism. However, in Examples 3 and 4 this will not be the case. If the true treatment assignment is ignorable, then $\Psi$ is still identified as in Equation \eqref{eq:gcomp}. To implement Equation \eqref{eq:gcomp} we need draws from the posterior of $\theta$ and $\eta$. Recall that the latter are the parameters governing the distribution of the outcome conditional on covariates $L$ and the \textit{true} exposure status, $A$, which we do not have. By Bayes' rule, this joint posterior over all unknowns, $ f(D^M, \xi, \eta, \gamma, \theta \mid D^O)$, is given by

\begin{equation} \label{eq:joint_misclass}
    \begin{split}
        f(a_1,\dots,a_n,\xi,  \eta, \gamma, \theta \mid D^O) \propto f(\xi) f_\Eta(\eta) f_\Gamma(\gamma)f_\Theta(\theta)\prod_{i=1}^n f_{\tilde A}(\tilde a_i\mid a_i, y_i, l_i; \xi) f_{Y|A,L}(y_i \mid a_i, l_i;\eta) f_{A|L}(a_i\mid l_i; \gamma) f_L(l_i;\theta)
    \end{split}
\end{equation}

Above, $f_{\tilde A}(\tilde a_i\mid a_i, y_i, l_i; \xi)=P(\tilde A=\tilde a_i\mid A=a_i, y_i, l_i; \xi)$ is the conditional probability mass function of the misclassified exposure evaluated at each subjects' observed value. Thus $f_{\tilde A}(\tilde a \mid a, y, l; \xi)$ is the misclassification model governed by sensitivity parameter $\xi$. Its specification is our first example of how sensitivity analysis is an ``art''. We can certainly specify this to have rich, elaborate dependence structures on $L$ and $Y$ governed by high dimensional $\xi$. These may allow us to explore more elaborate misclassification mechanisms, but at the expense of having many sensitivity parameters that are difficult to interpret. The ``art'' is in balancing interpretability with richness. As an example, a simple sensitivity analysis follows from exploring misclassified exposure distributions that are unrelated to (conditionally independent of) outcome and covariates so that $f_{\tilde A}(\tilde a \mid a, y,l; \xi)= f_{\tilde A}(\tilde a \mid a; \xi)$, which now essentially encodes the sensitivity (true positive) and specificity (true negative) of $\tilde A$. We let $\xi_1 = f_{\tilde A}(1 \mid 1; \xi)$ denote the sensitivity as it is the proportion of patients classified as exposed among those who were truly exposed. Similarly, we let $\xi_2 = f_{\tilde A}(1 \mid 0; \xi)$ denote 1 minus the specificity. Thus, the misclassification model's contribution to the posterior for each subject is 

$$f_{\tilde A}(\tilde a_i \mid a_i; \xi) = \xi_1^{\tilde a_i a_i} (1-\xi_1)^{(1-\tilde a_i) a_i} \xi_2^{\tilde a_i (1-a_i)} (1-\xi_2)^{(1-\tilde a_i) (1-a_i)} = \begin{cases}
  Ber(\tilde a_i; \xi_1), & a_i = 1, \\[6pt]
  Ber(\tilde a_i; \xi_2), & a_i = 0.
\end{cases}$$

The parameter vector $\xi=(\xi_1, \xi_2)$ is not identifiable from data since $D^O$ only contains the error-prone exposure indicator, $\tilde A_i$. A validation sub-sample with both $\tilde A_i$ and $A_i$ is typically not available and often prohibitively costly to collect. The case where $\xi_1=1$ and $\xi_2=0$ are the ``null'' values of the sensitivity parameters since it corresponds to 100\% sensitivity and specificity. This is the best-case scenario with no exposure mis-classification so that $\tilde A_i=A_i$ for each subject. On the other hand, $\xi_1 \approx 0$ and $\xi_2 \approx 1$ corresponds to a ``worse-case'' scenario in which sensitivity and specificity are nearly 0\%.

In terms of choices of priors, a common approach is to set the prior for the sensitivity parameter to a point-mass at a specified value. For instance, the priors $\xi_1 \sim \bar \delta_{1}(\xi_1)$ and $\xi_2 \sim \bar \delta_{0}(\xi_0)$ represent a strong prior belief in no exposure misclassification. Here $\bar\delta_{a}$ represents the point-mass distribution at value $a$. We can then conduct posterior inference under such point-mass priors across a range of non-null values to see how our posterior distribution over $\Psi$ changes. Under each prior, we obtain draws from the joint posterior $f(A_1=a_1,\dots, A_n=a_n,\xi,  \eta, \gamma, \theta \mid D^O)$, then, for each draw of $\eta$ and $\theta$, we evaluate the integral in Equation \eqref{eq:gcomp} as before.

The main impediment to implementation is that \stan's HMC sampler is gradient-based. Thus, it does not allow for discrete unknowns such as ($A_1=a_1, A_2=a_2, \dots, A_n=a_n$). One way to overcome this is to specify the log unnormalized \textit{marginal} posterior in which these discrete unknowns are integrated out. To that end, marginalizing out the $n$ unknown true exposure indicators from Equation \eqref{eq:joint_misclass} yields

\begin{equation*}
    \begin{split}
        f(\xi, \eta, \gamma, \theta \mid D^O) & = \sum_{a_1=0}^1 \sum_{a_2=0}^1...\sum_{a_n=0}^1 \int \int f(A_1=a_1,\dots, A_n=a_n, \xi, \eta, \gamma, \theta \mid D^O) \\
        & \propto f(\xi) f_\Eta(\eta) f_\Gamma(\gamma)f_\Theta(\theta)\prod_{i=1}^n \Big\{ \sum_{a_i=0}^1 f_{\tilde A}(\tilde a_i \mid a_i; \xi) f_{Y|A,L}(y_i \mid a_i, l_i) f_{A|L}(a_i\mid l_i; \gamma) \Big\} f_L(l_i;\theta)
    \end{split}
\end{equation*}

Now that the discrete unknowns are marginalized out, the remaining unknowns are continuous and so this joint marginal posterior can now be specified in \stan. The corresponding log unnormalized posterior joint density is 

\begin{equation*}
    \begin{split}
        \log \tilde f(\xi, \eta, \gamma, \theta \mid D^O) & =\log f(\xi) +\log f_\Eta(\eta) + \log f_\Gamma(\gamma)+\log f_\Theta(\theta) \\
        & \ \ \ \ \  + \sum_{i=1}^n \log\Big\{ \sum_{a_i=0}^1 f_{\tilde A}(\tilde a_i \mid a_i; \xi) f_{Y|A,L}(y_i \mid a_i, l_i) f_{A|L}(a_i\mid l_i; \gamma) \Big\} + \log f_L(l_i;\theta)
    \end{split}
\end{equation*}

Note that each subject's contribution involves a mixture of two components: $a_i=0$ and $a_i=1$. Thus, this is not a standard unnormalized posterior density and so must be specified using the \cd{target} incrementation as described in Section \ref{sc:custom}.

\begin{figure}[h!]
\begin{minipage}{.495\linewidth} \scriptsize
    \includegraphics[scale=.5]{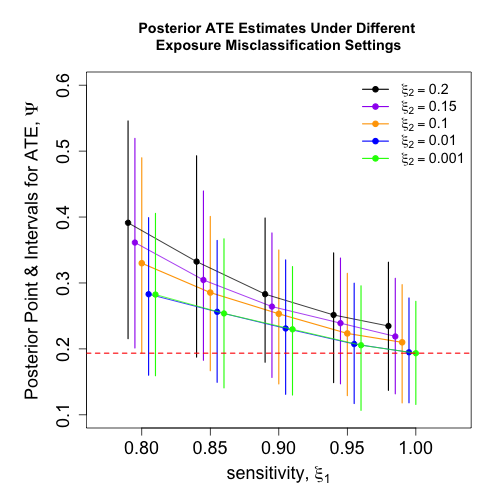}
\end{minipage}
\vline
\hfill
\begin{minipage}{.49\linewidth} 
    \includegraphics[scale=.15]{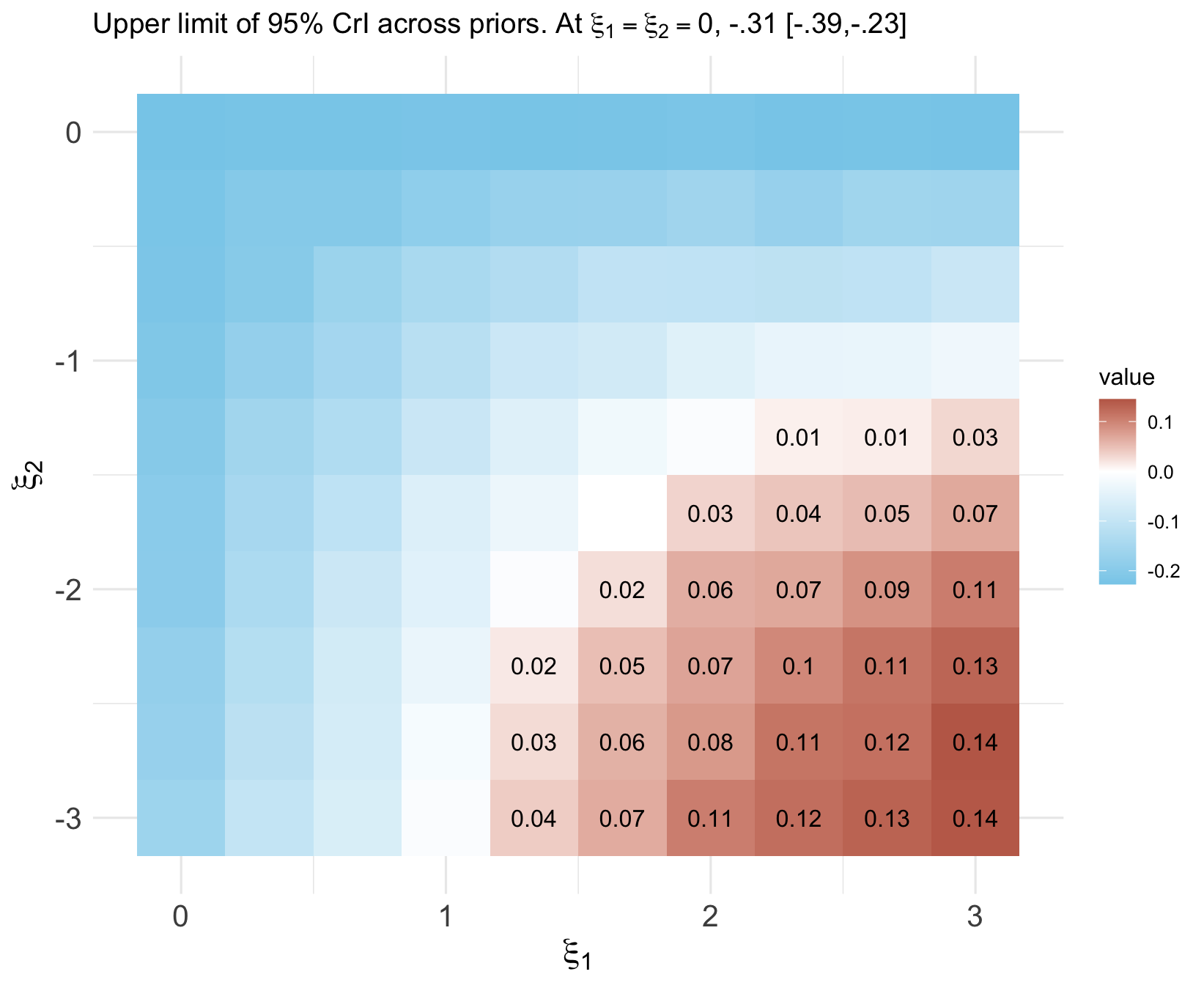}
\end{minipage}
\caption{ Left: posterior mean and 95\% credible intervals (CrI) for the ATE, $\Psi$, under various degrees of exposure misclassifications. The point at $\xi_1=.99$ on the green curve ($\xi_2=.001$) corresponds to the ``best case'' scenario where there is nearly no misclassification as sensitivity $=$ specificity $\approx 100\%$. The dashed red line represents the posterior mean of $\Psi$. In contrast, the point at $\xi_1=.8$ on the black curve $\xi_2=.2$ correspond to larger misclassification with sensitivity$=$specificity $\approx 80\%$. Right: tipping point analysis output from \stan. Assuming no unmeasured confounding, the posterior mean and 95\% CrI for $\Psi$ in the synthetic data analysis is -.31  [-.39, -.23], indicating lower remission rates due to $A=1$. At approximately values of $\xi_1>1$ and $\xi_2<-1$, however, the upper limit of the 95\% CrI for $\Psi$ surpasses zero. }
\label{fig:stan_misclass}
\end{figure}

As an example, the right panel of Figure \ref{fig:stan_misclass} visualizes posterior inference (mean and 95\% credible interval (CrI) ) for the ATE, $\Psi$. We use a synthetic dataset having $P=1$ binary confounder, $L$, a binary outcome, and point-mass priors for $\xi_1$ and $\xi_2$ at a grid of specified values. Accordingly, we specify outcome model $ f_{Y|A,L}(y \mid a, l) = Ber(y; \expit(\eta_0 +\eta_1a))$, treatment model $f_{A|L}(a\mid l; \gamma) = Ber(a; \expit(\gamma_0 +\gamma_1l))$, and covariate model $f_L(l;\theta) = Ber(l;\theta)$. Where the full parameter vectors for the outcome and treatment models respectively are $\eta=(\eta_0, \eta_1, \eta_2)$ and $\gamma = (\gamma_0, \gamma_1)$.The log unnormalized posterior joint density that was supplied to \stan for this analysis is 
\begin{equation*}
    \begin{split}
        \log \tilde f(\xi, \eta, \gamma, \theta \mid D^O) & = \log f_\Eta(\eta)  +\log f_\Gamma(\gamma)+\log f_\Theta(\theta) \\
        & \ \ \ \ \  + \sum_{i=1}^n \log\Big\{ \sum_{a_i=0}^1 Ber(\tilde a_i; \xi_{1}^{a_i}\xi_2^{1-a_i}) Ber(y_i; \expit((\eta_0 +\eta_1l_i + \eta_2a_i)) Ber(a_i; \expit((\gamma_0 +\gamma_1l_i)) \Big\} \\
        & \ \ \ \ \  + \log Ber(l_i;\theta)
    \end{split}
\end{equation*}

Since manual coding of the middle mixture term in \stan is often numerically unstable, we recommend using the more stable built-in function \cd{log\_sum\_exp()} when evaluating it during the \cd{target} incrementation. As an example, here is the relevant excerpt of $\cd{exposure\_misclassification.stan}$ which does the middle mixture term. The full code can be found on the GitHub repository and the ellipsis denotes non-essential lines skipped here for ease of exposition.

\begin{lstlisting}[title={},captionpos=t, caption={},numbers=none]
model {
  
  ...  
  // mixture term likelihood contribution
  for(i in 1:n){
    target += log_sum_exp( 
    
    // mixture term for a=1
    bernoulli_lpmf(a_tilde[i] | xi1) 
    + bernoulli_lpmf(y[i]|inv_logit(eta[1]+eta[2]*l[i]+eta[3]))
    + bernoulli_lpmf(1 | inv_logit(gamma[1] + gamma[2]*l[i])),   
    
    // mixture term for a=0
    bernoulli_lpmf(a_tilde[i] | xi2)
    + bernoulli_lpmf(y[i] | inv_logit( eta[1] + eta[2]*l[i]))
    + bernoulli_lpmf(0 | inv_logit( gamma[1] + gamma[2]*l[i])) 
     );
   } 
}
\end{lstlisting}

Using \stan, we conducted posterior inference under a grid of point-mass priors for the two sensitivity parameters. The point at $\xi_1=.99$ on the green curve ($\xi_2=.001$) corresponds to estimates close to the ``null value'' of the sensitivity parameter. That is, this is the ``best case'' scenario where there is nearly no misclassification since sensitivity $=$ specificity $\approx 100\%$. The dashed red line represents the posterior mean of $\Psi$ under these priors. In contrast, the point at $\xi_1=.8$ on the black curve $\xi_2=.2$ is the estimate under ``non-null'' values of the sensitivity parameter. Here, there is larger  misclassification bias with sensitivity$=$specificity $\approx 80\%$. Notice that at this point, the 95\% CrI excludes the ``best case'' point estimate. Thus this $80\%$ value corresponds to a kind of ``tipping point''  - i.e. the degree of misclassification bias that would overturn posterior inferences made under the ``null value'' of the sensitivity parameters. In \stan, such point-mass priors are set by supplied the values in the \cd{data} block - essentially treating them as fixed at those values.

Finally, we note that while we made a simplifying assumption in the specification of the misclassification model that $f_{\tilde A}(\tilde a \mid a, y,l; \xi)= f_{\tilde A}(\tilde a \mid a; \xi)$, this is not necessary. We could, for instance, allow sensitivity and specificity to depend on levels of the covariate, $L$. Indeed, if we have prior information that misclassification depends on say, sex, we should consider this. However, this comes at the cost of introducing more sensitivity parameters, which complicates interpretation. In this way, all sensitivity analysis is an exercise in balancing simplicity while allowing for exploration of sufficiently rich and realistic assumption violations. This balance is inherently application and domain-specific.

\subsection{Example 2: Unmeasured Confounding} \label{sc:unmeasured confounding}

In this setting, we consider another example of a Bayesian sensitivity analysis in \stan around the ``no unmeasured confounding'' assumption which asserts that $A\perp Y^1,Y^0 \mid L$. In some scenarios, we may not believe this assumption holds as subject-matter expertise may inform us that a plausible unmeasured covariate, $U$, which impacted both selection into one of the treatment groups, $A$, as well as the potential outcomes - was left unrecorded in the data. That is, while we think that $A \not \perp Y^1, Y^0  \mid L$, we believe that $A \perp  Y^1, Y^0 \mid L, U$. Hereafter, for simplicity of exposition we assume $U\sim N(0,1)$ is a standardized continuous marker so that we can interpret results in unit-free terms via standard deviations. We denote the standard normal density function with $\phi(u)$.

We also conservatively assume that $U\perp L$ - that is, this unmeasured confounder is unrelated to the observed covariates we adjust for, $L$. To see why this is conservative, consider the extreme case where $U$ is deterministically related to one of the covariates in $L$. Then, adjusting for this component of $L$ is the same as adjusting for $U$ and the unmeasured confounding concern dissipates. Under this modified ignorability assumption if SUTVA and positivity hold, $\Psi$ can be identified as 

\begin{equation}
    \Psi(\eta, \theta, \xi) = \int\int \Big( E[Y\mid A=1, L=l, U=u;\eta,, \xi] - E[Y\mid A=0, L=l, U=u;\eta,\xi ]\Big) \cdot f_{L}(l;\theta)\phi(u) \ dl \ du
\end{equation}

Note that unlike before, the causal estimand is a function of non-identifiable sensitivity parameters, $\xi$ - components of which may govern the outcome model's dependence on the unmeasured confounder, $U$. Intuitively, this parameter is non-identifiable because $U$ is never observed and so the dependence between $U$ and $Y$ at a given level of $A$ and $L$ cannot be informed by the observed data. However, the Bayesian treatment of the problem is the same as in the misclassification case. Here, $D^O=\{\delta_i=1, y_i, a_i, l_i\}$ and $D^M=\{U_i\}_{i=1}^n$. As before, $\delta_i$ is an indicator with 1 indicating that $U_i$ is missing. In ``unmeasured confounding'' settings, this indicator $\delta_i=1$ for all $i$ and so the missingness mechanism can again be ignored. The missing data, $D^M$, contains the $n$ unknown confounder values for the subjects in our sample. As before, we must now draw from the joint posterior over  all unknowns, $f(D^M, \xi, \eta, \gamma, \theta \mid D^O)$, is 

\begin{equation*}
    f( u_1, u_2,\dots, u_n, \xi, \eta, \gamma, \theta \mid D^O) \propto f(\xi)f_\Eta(\eta) f_\Gamma(\gamma) f_\Theta(\theta) \prod_{i=1}^n f_{Y|A,L,U}(y_i\mid a_i, l_i, u_i; \eta, \xi) f_{A|L,U}(a_i\mid l_i, u_i; \gamma, \xi)f_L(l_i; \theta) \phi(u_i)
\end{equation*}

Note that following Bayes' rule we see that $U$ is in the conditional of both the outcome and the treatment model. This should make intuitive sense as a covariate is a confounder only if it impacts both outcome and treatment. Just as with misclassification we had to make simplifying assumptions about the misclassification via the conditional pmf $f_{\tilde A}(\tilde a \mid a, y,l; \xi)$, here we must make simplifying assumptions about the dependence of outcome and treatment on $U$. As in the misclassification case, this is an application-specific art that balances interpretability against the complexity of violations that are explored in the sensitivity analysis. For concreteness, suppose we have a single binary observed confounder, $L$, and a binary outcome, $Y\in\{0,1\}$. Then we could specify

\begin{equation*}
    \begin{split}
        P(Y=1\mid a, l,  u; \eta, \xi_1) & = \expit( \eta_0 +\eta_1a + \eta_2l + \xi_1 u ) \\
        P(A=1 \mid l, u; \gamma, \xi_2) & = \expit(\gamma_0 + \gamma_1 l + \xi_2 u )
    \end{split}
\end{equation*}

Here $\eta=(\eta_0, \eta_1, \eta_2)$ and $\gamma=(\gamma_0, \gamma_1)$. This encodes a linear dependence (on the logit-scale) between the unmeasured confounder with both treatment and outcome probability. For example, it asserts that, say, among treated patient with covariates $L=l$, the odds of the outcome among the subset with a one-standard deviation higher than average $U$ is $\exp(\xi_1)$ times the odds of the outcome among those with average $U$.

A strong prior belief in ``no unmeasured confounding'' is encoded by either setting $\xi_1 \sim \bar \delta_{0}(\xi_1)$ for any $\xi_2 \in \mathbb{R}$, or setting $\xi_2 \sim \bar \delta_{0}(\xi_2)$ for any $\xi_1 \in \mathbb{R}$, or setting both $\xi_1 \sim \bar \delta_{0}(\xi_1)$ and $\xi_2 \sim \bar \delta_{0}(\xi_2)$. Thus, these define the ``null values.'' Point-mass priors at non-null values encode \textit{departures} from the assumption with the magnitude and sign of the parameters indicating the strength and direction of the departure. ``Reasonable'' non-null values may be determined from published literature or subject-matter expertise. Using \stan we can trace out posterior inferences across a grid of sensitivity parameter values to explore how results change as we depart from null values. An example, which we will discuss subsequently, is in the right panel of Figure \ref{fig:stan_misclass}. Given the binary outcome and treatment models above and point mass priors on specified sensitivity parameters, the corresponding log unnormalized posterior density to be specified in \stan is 

\begin{equation*}
    \begin{split}
            \log \tilde f(  u_1, u_2,\dots, u_n, \xi, \eta, \gamma, \theta \mid D^O) & = \log f(\xi) + \log f_\Eta(\eta) + \log f_\Gamma(\gamma) +\log f_\Theta(\theta) \\
            & \ \ \ \ + \sum_{i=1}^n \log Ber(y_i;  \expit( \eta_0 +\eta_1a_i + \eta_2l_i + \xi_1 u_i ) ) + \log Ber(a_i; \expit(\gamma_0 + \gamma_1 l_i + \xi_2 u_i )) \\
            & \ \ \ \ + \log Ber(l_i; \theta) + \log \phi(u_i)
    \end{split}
\end{equation*}

Unlike in the measurement error example which involved discrete unknowns, here all unknowns including $U$ are continuous. Thus analytic marginalization is not required to sample from this posterior in \stan. Instead, in order to obtain draws from the correct joint posterior corresponding to $ \log \tilde f( u_1, u_2, \dots, u_n, \xi, \eta, \gamma, \theta \mid D^O)$ above, the unobserved $U_i$ should be declared as a ``parameter'' in the parameter block and given a standard normal ``prior'' in the model block of the \stan file, as shown in the following excerpt from \cd{unmeasured\_mconfounding.stan}:

\begin{lstlisting}[title={},captionpos=t,caption={},numbers=none,
                   label=labbb]
parameters {
  ... 
  real u[n];
  ...
}

model {
 ...
 u ~ normal(0,1);
 ...

 for(i in 1:n){
  y[i] ~ bernoulli(inv_logit(eta0+eta1*a[i]+eta2*l[i]+ xi1*u[i]));
  a[i] ~ bernoulli(inv_logit(gam0 + gam1*l[i] + xi2*u[i]));
  l[i] ~ bernoulli(theta1);
 } 
}
\end{lstlisting}

As an example, we implement the method on a synthetic data of $n=300$ with a binary outcome and a single measured binary covariate, $L$, as described above. Concretely, suppose that $Y\in\{0,1\}$ is an indicator of remission within 3 years of treatment start and we would like to contrast 3-year remission rate between treatment $A=1$ and $A=0$ while adjusting for $L$. Running the models above with strong priors on no unmeasured confounding $\xi_1, \xi_2 \sim \bar \delta_0$ yields a posterior mean and credible of $-.31, 95\%$ CrI:$[-.39, -.23]$ for $\Psi$. This seems to indicate lower remission rates for treatment $A=1$.

Now, suppose upon further consultation with subject-matter experts we are concerned that baseline toxicity level, $U$, is a confounder that has gone unmeasured. Thus while treatment $A=1$ may seem to lead to lower remission, it may be that subjects with higher toxicity were more likely to get $A=1$ ($\xi_2>0$), and less likely to have remission ($\xi_1 <0$). Then we cannot determine whether lower remission rate was due to $A=1$ or due to treated units having higher toxicity.

In this case, since our findings are in favor of treatment $A=1$, a reasonable question to ask is how strong unmeasured confounding with this specified structure would need to be before our posterior beliefs (as measured by a 95\% of posterior probability mass) includes $\Psi=0$. To that end, the heat map in the right panel of Figure \ref{fig:stan_misclass} displays the upper limit of the 95\% CrI across a grid of values for $\xi_1<0$ and $\xi_2>0$. We see in the figure that this occurs approximately at $(\xi_1 \approx 1, \xi_2 \approx -1)$. That is, if - treatment and measured covariates $L$ being the same - subjects with a 1-standard deviation higher than average toxicity have a $\exp(1)\approx 2.7$ times higher odds of remission relative to patients with average toxicity whiled having $\exp(-1) \approx .36$ times lower odds of getting treatment $A=1$. Log odds ratios around 1 and or -1 are small and fairly typical and thus we may conclude that our results are sensitive to unmeasured confounding as only a small departure from the assumption would shift our results. Such analyses are sometimes called ``tipping point'' analyses but it may be useful to present the entire heatmap rather than just the tipping point thresholds of $(\xi_1 \approx 1, \xi_2 \approx -1)$ as in the right panel of Figure \ref{fig:stan_misclass}.

\subsection{Example 3: Outcomes Missing Not-at-Random} \label{sc:ex_mnar_parametric}

In many causal inference settings, the outcome of interest may only be observed for a subset of the sample. Suppose, as in Example 2, we are interested in estimating the effect of a binary treatment on a binary outcome. As a concrete example we will again say this is remission within 3-years of treatment initiation, $Y\in\{0,1\}$. Suppose that exposures are accurately classified and the treatment assignment mechanism is strongly ignorable. For various reasons, we may not have the full three years of information required to accurately ascertain $Y$ for all subjects. For instance, in electronic health records, patients in the sample may remain remission-free before transferring out of the health system for which you have records before 3 years. In insurance claims data bases, patients remain remission-free but then lose insurance eligibility before the 3-year mark. Even in well-designed clinical trials, patients may remission-free and be withdrawn from the study before 3-years due to toxicity. In all these cases, we do not know whether such patients would have gone on to achieve remission within three years. That is, $Y$ is missing.

In such problems, a typical observed data structure is $D^O = \{\delta_i, y_i, a_i, l_i\}$, where $\delta_i=1$ if $Y_i$ is missing and $\delta_i=0$ if it is observed. We let $n_m=\sum_{i=1}^n \delta_i$ be the number of subjects with missing outcomes and $n_o=n-n_m$ be the number of patients with observed outcomes. Without loss of generality, we assume that the data are ordered so that the $n_o$ patients with observed outcomes appear first in the data with index running $i = 1,2,\dots, n_o, n_o+1, n_o+2, \dots, n_o +n_m =n$. So, for example we can equivalently write $D^O = \{ \delta_i=0 ,y_i, a_i l_i\}_{i=1}^{n_o} \cup \{  \delta_i=1, a_i, l_i \}_{i=n_o+1}^n$. In this problem, the missing data $D^M = \{y_i\}_{i=n_o+1}^n$ is comprised of the outcome values for the $n_m$ subjects who have it missing. The complete data we ideally would observe is $D^C = D^O \cup D^M$. Note that, unlike in Examples 1 and 2 in which  we had $\delta_i=1$ for all subjects regardless of their observed or missing data, here $\delta_i=1$ only for some subjects. Thus, we must explicitly model its conditional distribution - i.e. the missingness mechanism, $P(\Delta =1 \mid Y, A, L;\xi)$.

In such settings, the outcome model required for the g-formula in Equation \eqref{eq:gcomp} may be estimated using the subset of complete-case subjects (i.e. those with $\delta_i=0$) under a missing-at-random (MAR) assumption\cite{daniels2008missing, molenberghs2014}. This is an assumption about the missingness mechanism and asserts $\Delta \perp Y \mid A, L$. In other words, a subject whose outcome is missing is exchangeable with a subject whose outcome is observed as long as they have the same treatment and covariate values. Note that unlike in standard missing-data settings, here the missingness mechanism involves the treatment of interest and thus must be thought about jointly with the treatment mechanism - a unique consequence of the causal framework the missingness is embedded within.

In many settings we may believe that missingness is not at-random (MNAR), $\Delta \not \perp Y \mid A, L$. In the clinical trial example, for instance, if withdrawal was due to toxicity we may believe that patients with missing outcomes were less likely to have remission relative to patients who were not withdrawn, even if they had the same covariate values $L=l$ and were in the same treatment arm $A=a$.

In this case, the joint posterior $ f(D^M, \xi, \eta, \gamma, \theta \mid D^O)$ is given by 

\begin{equation*}
    \begin{split}
        f(y_{n_o+1}, y_{n_o+2}, \dots, y_{n}, \xi, \eta, \gamma, \theta \mid D^O) & \propto  f(\xi)f_\Eta(\eta)f_\Gamma(\gamma)f_\Theta(\theta) \\
        & \ \ \ \ \times \prod_{i=1}^{n_o} P(\Delta =0 \mid y_i, a_i, l_i;\xi) f_{Y|A,L}(y_i \mid a_i,l_i;\eta) f_{A|L}(a_i \mid l_i;\gamma) f_L(l_i;\theta) \\
        & \ \ \ \ \times \prod_{i=n_o+1}^n P(\Delta =1 \mid y_i, a_i, l_i;\xi) f_{Y|A,L}(y_i \mid a_i,l_i;\eta) f_{A|L}(a_i \mid l_i;\gamma) f_L(l_i;\theta)
    \end{split}
\end{equation*}

Under MAR, the missingness mechansim becomes $P(\Delta =1 \mid Y, A, L;\xi)=P(\Delta =1 \mid A, L;\xi)$. So, to perform the g-computation in Equation \eqref{eq:gcomp}, it is sufficient sample from the marginal posterior

\begin{equation*}
    \begin{split}
        f(\eta,\theta\mid D^O) & = \sum_{y_{n_o+1}=0}^1 \sum_{y_{n_o+2}=0}^1 ...\sum_{y_{n}=0}^1 \int \int  f(y_{n_o+1}, y_{n_o+2}, \dots, y_{n}, \xi, \eta, \gamma, \theta \mid D^O) d\xi d\gamma \\
                              & = C \times \prod_{i=1}^{n_o} f_{Y|A,L}(y_i \mid a_i,l_i;\eta) f_L(l_i;\theta) \prod_{i=n_o+1}^n f_L(l_i;\theta)\\
    \end{split}
\end{equation*}
Here, the multiplicative constant, $C$, is 
$$ C = \int \int \prod_{i=1}^{n} P(\Delta =\delta_i \mid a_i, l_i;\xi) f_{A|L}(a_i \mid l_i;\gamma) f(\xi) f_\Gamma(\gamma) d\xi d\gamma $$

and can be absorbed into the proportionality constant since it does not involve the two parameters of interest, $(\eta,\theta)$. Note that in the marginal posterior under MAR, only patients with observed outcomes contribute information for $\eta$, while all patients contribute information about $\theta$.

In MNAR settings, the process is more complicated since the missingness mechanism cannot be absorbed into the proportionality constant and the missing outcomes cannot be marginalized out as above. So, as in the previous examples, we instead sample from the joint posterior of all unknowns, $f(y_{n_o+1}, y_{n_o+2}, \dots, y_{n}, \xi, \eta, \gamma, \theta \mid D^O)$. In this case, this involves modeling the missingness mechanism directly via sensitivity parameters. As before, there are many valid, application-specific ways to do this. We will give one example in which we have a binary outcome, a single $P=1$ binary observed covariate, $L$, and specifying the missingness mechanism to be 

$$  P(\Delta =1 \mid y_i, a_i, l_i;\xi) = \expit\big ( \xi_0 + \xi_1a+\xi_2y + \xi_3 (a\cdot y) + \xi_4l \big )$$

Note that the key sensitivity parameters here are $\xi_2$ and $\xi_3$. This model asserts that among untreated subjects with covariates $L=l$ patients with remission have an odds of missingness that is $\exp(\xi_2)$ times the odds of missingness among those without remission. Similarly, it asserts that among treated patients with covariates $L=l$, patients with remission have an odds of missingness that is $\exp(\xi_2+\xi_3)$ times the odds of of missingness among those without remission. The ``null values'' which encode MAR are therefore $\xi_2=\xi_3=0$. Other values are ``non-null'' in that they encode departures from MAR, with magnitude and sign of the parameters indicating the degree and direction of the resulting bias, respectively. Note that it assumes the odds ratio of the observed covariates $L$ on $\Delta$ is the same across treatment and remission levels. Again, this is in the spirit of balancing simplicity with richness.

Regarding specification of the posterior in \stan, notice that the joint posterior distribution includes the missing binary outcomes for the $n_m$ subjects. If the outcome was continuous, we could declare them in the \cd{parameters} block of the \cd{.stan} file as we did in the unmeasured confounding example. However, as mentioned in the misclassification example, discrete unknowns are not allowable in the\cd{parameters} block. And so we must once again supply \stan with a marginal posterior in which these discrete parameters are marginalized out. This posterior is 

\begin{equation*}
    \begin{split}
        f(\xi, \eta,\theta \mid D^O) & = \sum_{y_{n_o+1}=0}^1 \sum_{y_{n_o+2}=0}^1 ...\sum_{y_{n}=0}^1 \int f(y_{n_o+1}, y_{n_o+2}, \dots, y_{n}, \xi, \eta, \gamma, \theta \mid D^O) d\gamma \\
                              & \propto  f(\xi) f_\Eta(\eta) f_\Theta(\theta)  \prod_{i=1}^{n_o} P(\Delta =0 \mid y_i, a_i, l_i;\xi) f_{Y|A,L}(y_i \mid a_i,l_i;\eta) f_L(l_i;\theta) \\
        & \ \ \ \ \times \prod_{i=n_o+1}^n \Big\{ \sum_{y_i=0}^1 P(\Delta =1 \mid y_i, a_i, l_i;\xi) f_{Y|A,L}(y_i \mid a_i,l_i;\eta)  \Big\}f_L(l_i;\theta)
    \end{split}
\end{equation*}

Note here that because the g-formula does not involve the treatment model parameters $\gamma$ and the treatment model does not involve the missing data, we can drop the treatment model integrated over $\gamma$ while maintaining proportionality. The corresponding log unnormalized posterior density that must be supplied to \stan is then 
\begin{equation*}
    \begin{split}
        \log \tilde f(\xi, \eta,\theta \mid D^O) & =  \log f(\xi) + \log f_\Eta(\eta) + \log f_\Theta(\theta) \\
        & \ \ \ \ \ + \sum_{i=1}^{n_o} \log P(\Delta =0 \mid y_i, a_i, l_i;\xi) + \log f_{Y|A,L}(y_i \mid a_i,l_i;\eta) +\log f_L(l_i;\theta) \\
        & \ \ \ \ \times \sum_{i=n_o+1}^n \log \Big\{ \sum_{y_i=0}^1 P(\Delta =1 \mid y_i, a_i, l_i;\xi) f_{Y|A,L}(y_i \mid a_i,l_i;\eta)  \Big\}+\log f_L(l_i;\theta)
    \end{split}
\end{equation*}

As in the misclassification example, given the two-component mixture contribution of subjects with missing outcome information in the third line of the equation above, we recommend specification using the \cd{target} syntax with built-in functions like \cd{log\_exp\_sum()} for numerical stability.

\begin{figure}[h!] 
    \centering
    \includegraphics[scale=.45]{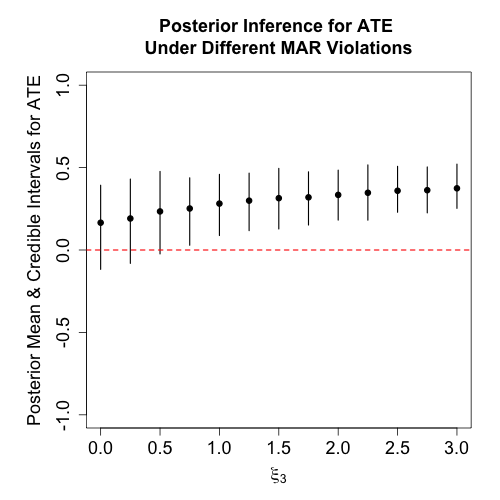}
\caption{ Results from synthetic example described in Section \ref{sc:ex_mnar_parametric}. A value of $\xi_3=0$ represents strong prior belief in MAR - leading to a credible interval for $\Psi$ that contains zero. However, as we allow deviations from MAR, posterior inferences shift in favor of a positive effect for group $A=1$ versus $A=0$ at about $\xi_3=.75$.}
\label{fig:stan_mnar}
\end{figure}

In order to fix ideas, we present another synthetic example with $n=1000$ subjects in which only $n_o=74$ subjects had an observed outcome. The outcome is binary and we have a single $P=1$ binary measured covariate, $L\in\{0,1\}$. Along with the missingness mechanism model above, we specify the following logistic outcome model

$$ f_{Y|A,L}(y \mid a,l;\eta) = Ber(y \mid \expit( \eta_0+\eta_1l + \eta_2a) ) $$

A standard analysis under MAR corresponds to $\xi_2,\xi_3 \sim \bar \delta_0$. This yields a posterior mean and 95\% CrI for $\Psi$ of $0.17, [-0.09, 0.41 ]$. Thus, we see the posterior is centered around a higher effect on remission risk for treatment $A=1$ relative to $A=0$ with large uncertainty as reflected with the wide CrI.

Suppose, however, that after conversations with clinical experts we think that subjects with $A=0$ who have outcomes observed are representative of that subpopulation. However, treated patients with missing outcomes were more likely to have gone on to attain remission. Thus, in the sample of treated subjects with observed data the proportion of remission would be lower than in the target population. This would shift our posterior under MAR towards $\Psi \approx 0$. This belief can be encoded via the following missingness mechanism specification

$$  P(\Delta =1 \mid y, a, l;\xi) = \expit\big ( \xi_0 + \xi_1a + \xi_3 (a\cdot y) \big )$$

The above model 1) allows for different proportions of missingness in treatment group, 2) asserts that $\Delta \perp Y \mid A=0, L$, and 3) asserts $\Delta \not \perp Y \mid A=1, L$. The last assertion is encoded by $\xi_3\neq 0$. In particular, the information we gained from our clinical expert suggests that $\xi_3>0$, since treated patients with remission are more likely to have that outcome missing. The right panel of Figure \ref{fig:stan_mnar} shows how posterior inference for $\Psi$ changes across a grid of prior belief about this value. At $\xi_3=.75$ ($\exp(.75)\approx 2)$, we see that the posterior interval excludes $\psi=0$. Given the large proportion of subjects with missing outcomes, it is not surprising that such a low log-odds ratio is sufficient to shift 95\% of the posterior probability mass away from $\Psi=0$. We may conclude, therefore, that our results are sensitive to this type MAR violation. However, since whether the OR is small enough to be deemed ``sensitive'' is subjective, in general it is preferable to display the entire curve over a grid as in Figure \ref{fig:stan_mnar}. The log unnormalized posterior above for this specific problem is
\begin{equation*}
    \begin{split}
        \log \tilde f(\xi, \eta,\theta \mid D^O) & =  \log f(\xi) + \log f_\Eta(\eta) + \log f_\Theta(\theta) \\
        & \ \ \ \ \ + \sum_{i=1}^{n_o} \log Ber\big(0; \expit ( \xi_0 + \xi_1a_i + \xi_3 (a_i\cdot y_i) )\big) + \log Ber(y_i \mid \expit( \eta_0+\eta_1l_i + \eta_2a_i) ) +\log Ber(l_i; \theta) \\
        & \ \ \ \ \times \sum_{i=n_o+1}^n \log \Big\{ \sum_{y_i=0}^1 Ber\big(1; \expit ( \xi_0 + \xi_1a_i + \xi_3 (a_i\cdot y_i)  ) \big) Ber(y_i \mid \expit( \eta_0+\eta_1l_i + \eta_2a_i) ) +\log Ber(l_i; \theta) \Big\}
    \end{split}
\end{equation*}

Here is the relevant excerpt in from \cd{mnar\_outcome.stan} in which we program the two-component mixture contribution in the last line:

\begin{lstlisting}[title={},captionpos=t,caption={},numbers=none]
data{
  ...  
  int a_m[n_m];
  int l_m[n_m];
  ...
}
...
model{  
  ...  
  for(i in 1:n_m){    
    target += log_sum_exp( 
         bernoulli_lpmf(1 | inv_logit(xi[1] + xi[2]*a_m[i] + xi2 + xi3*a_m[i])) 
         + bernoulli_lpmf(1 | inv_logit( eta[1] + eta[2]*l_m[i] + eta[3]*a_m[i])),
                    
         bernoulli_lpmf(1 | inv_logit(xi[1] + xi[2]*a_m[i])) 
         + bernoulli_lpmf(0|inv_logit(eta[1]+eta[2]*l_m[i]+eta[3]*a_m[i]))
    );            
  } 
}
\end{lstlisting}

Once draws of $\eta$ and $\theta$ are obtained from \stan, Equation \eqref{eq:gcomp} is evaluated with each draw to obtain a posterior draw of $\Psi$. Since the procedure and implementation in the \cd{generated quantities} block is the same as in Section \ref{sc:bayesian_causal_standard}, we do not describe it again here.

\section{Example 4: Bayesian Nonparametric Sensitivity Analyses} \label{sc:ex_mnar_tsb}

While the previous examples used simple, low-dimensional models to keep focus on sensitivity analysis and implementation concepts, a major strength of the Bayesian approach to causal inference is the availability of flexible (i.e. high-dimensional) models. One common class of flexible Bayesian models are truncated stick-breaking (TSB) mixtures \cite{Ishwaran2001,Ishwaran2002}. The joint distribution of the observed outcome, treatment, and covariates are modeled as a finite mixture
\begin{equation} \label{eq:tsb_joint}
    f(y,a, l; \omega, \bm v) = \sum_{k=1}^K \nu_k f_{Y|AL}(y \mid a, l; \eta_k)f_{A|L}(a \mid l; \gamma_k)f_L(l; \theta_k)
\end{equation}

Where $\nu=\{ \nu_k\}_{k=1}^K$ are the mixture weights - note these are Greek ``nu'' which are distinct from the latin $V$ to be introduced subsequently. Here, $\nu$ is a vector in the $n$-simplex - the space of all length-$n$ vectors with $\nu_k>0$ for each $k$ and $\sum_k \nu_k =1$. We also define $\omega = \{\omega_k\}_{k=1}^K$ with each $\omega_k = (\eta_k,\gamma_k, \theta_k)$ being the mixture component parameters. We will denote the factor-specific collection of parameters with $\eta=\{\eta_k\}_{k=1}^K$, $\gamma=\{\gamma_k\}_{k=1}^K$, and $\theta=\{\theta_k\}_{k=1}^K$. The component parameters are typically assigned a joint prior distribution (sometimes called a ``base distribution'' in the TSB mixture literature) with density $g_{\Omega}(\omega)$. In a TSB mixture, the weights are generated as follows: for each $k=1,2,\dots, K-1$, we set 
$$\nu_k = V_k \prod_{j=1}^{k-1} (1 - V_j)$$ 
with $\nu_K = \prod_{j=1}^{K-1} (1 - V_j)$. Here, the $\bm V=(V_1,V_2,\dots, V_{K-1})$ are independent Beta random variables, $V_k \sim Beta(1,\alpha)$, for $k=1,\dots,K-1$. So the prior density is $f_{\bm V}(\bm v) = \prod_{k=1}^{K-1} Beta(v_k; 1,\alpha)$. That is, the $\nu$ parameters are just transformations of $\bm V$, and so the independent Beta priors on the latter induce a prior on the former. The parameter $\alpha > 0$ is known as the concentration parameter and is often given a $Gam(1,1)$ hyperprior\cite{Escobar1995}. This is known as ``stick-breaking'' because each $\nu_k$ is obtained by ``breaking off'' some portion, $V_k$, of the remaining segment, $\prod_{j=1}^{k-1} (1 - V_j)$, of a unit-segment (i.e. the ``stick''). The stick-breaking prior on the weights prevents overfitting by encouraging few, large weights rather than, say, uniform weights across the $K$ components.

$\bm V$ along with the component parameters $\omega=(\eta, \gamma, \theta)$ represent the unknowns in this model, while $f_{\bm V}(\bm v)$ and the base distribution, $g_\Omega(\omega)$, represent the priors. Even if the component models are specified to be parametric, the induced model - marginal of the mixture components - are flexible. For instance, the induced marginal outcome regression is a mixture of the component-specific outcome regressions

\begin{equation} \label{eq:tsb_outcome}
    E_{Y|AL}[Y\mid A=a, L=l; \omega, \bm v] = \sum_{k=1}^K w_k(a,l; \gamma,\theta,\bm v) E_{Y|AL}[Y \mid A=a, L=l; \eta_k]
\end{equation}
with treatment/covariate-dependent weights
$$ w_k(a,l; \gamma,\theta, \bm v) = \frac{ \nu_k f_{A|L}(a\mid l;\gamma_{k})f_L(l;\theta_{k}) }{\sum_{k'=1}^K \nu_{k'} f_{A|L}(a\mid l;\gamma_{k'})f_L(l;\theta_{k'}) }$$

That is, an outcome regression local to a particular mixture component gets higher weight if the parameters of the treatment and covariate models in that mixture component yield a higher density/mass evaluation at point $(a,l)$. Mixture components with treatment/covariate model parameters less consistent with a given point $(a,l)$ get less weight. In summary, the induced outcome regression is a flexible treatment-covariate dependent mixture of parametric regressions. Figure \ref{fig:tsb_mnar} gives examples of the regression lines relating $Y$ and $L$ for each treatment group, $A\in\{0,1\}$ to give some intuition about the flexability. Similarly, the induce marginal distribution of the covariates is given by 

\begin{equation} \label{eq:tsb_covariate}
    f_L(l;\theta,\bm v) = \sum_{k=1}^K \nu_k f_L(l;\theta_k)
\end{equation}

The choice of component models is application specific and will partially depend on the support of the outcome and covariates (e.g. discrete versus continuous). In the general case, for a particular specification of each mixture component models $f_{Y|AL}(y \mid a, l; \eta_k)$, $f_{A|L}(a \mid l; \gamma_k)$, and $f_L(l; \theta_k)$, along with prior $g_\Omega(\omega)$, the log unnormalized posterior density is 

$$ \log \tilde f(\bm v, \omega \mid D^o) \propto \log g_\Omega(\omega) +\log f_{\bm V}(\bm v)+ \sum_{i=1}^n \log\Big( \sum_{k=1}^K \nu_k f_{Y|AL}(y_i \mid a_i, l_i; \eta_k)f_{A|L}(a_i \mid l_i; \gamma_k)f_L(l_i; \theta_k)\Big)$$

This can be specified in the \cd{model} block of a \stan file to obtain the $m^{th}$ posterior draw of the component parameters $\omega_k^{(m)} = \{ \eta_k^{(m)}, \gamma_k^{(m)}, \theta_k^{(m)}\}_{k=1}^K$ and a draw of the stick-breaking weights $\{\bm v_k^{(m)}\}_{k=1}^K$. To compute the $m^{th}$ posterior draw of the ATE, plug these parameters into Equations \eqref{eq:tsb_outcome} and \eqref{eq:tsb_covariate} and standardize according to the formula in \eqref{eq:gcomp}:
$$ \Psi^{(m)} = \int \Big(E[Y\mid A=1, L=l;\omega^{(m)},\bm v^{(m)}] - E[Y\mid A=0, L=l;\omega^{(m)},\bm v^{(m)}]\Big)f_L(l;\theta^{(m)},\bm v^{(m)}) dl$$
This can be done in the \cd{generated quantities} block in \stan as we have done previously. The evaluation of the integral will typically require monte carlo simulations from $f_L(l;\theta^{(m)},\bm v^{(m)})$ for each posterior parameter draw $m$, which can also be done in the \cd{generated quantities} block.

\subsection{Extension to MNAR Outcomes}

Using the same notation as in Section \ref{sc:ex_mnar_parametric}, we now consider the case with missing outcomes in which $D^O = \{\delta_i, y_i, a_i, l_i\}$ and $D^M = \{y_i\}_{i=n_o+1}^n$. Without assuming that the outcome missingness is at-random, we must model the missingness mechanism as Bernoulli with conditional probability $P(\Delta=1\mid y, a, l;\xi) $ explicitly once again. Under the TSB model for the joint distribution presented in \eqref{eq:tsb_joint}, the joint posterior over all unknowns - including the missing data and sensitivity parameters - is proportional to

\begin{equation} \label{eq:mixture_post_missing}
    \begin{split}
        f(D^M,\xi,\omega,\bm v \mid D^O) & \propto g_\Omega(\omega)f_{\bm V}(\bm v) \prod_{i=1}^n P(\Delta=\delta_i\mid y_i, a_i,l_i;\xi) f(y_i, a_i, l_i; \omega, \bm v)
    \end{split}
\end{equation}
where each $f(y_i, a_i, l_i; \omega, \bm v)$ has the mixture form in Equation \ref{eq:tsb_joint}. This is perhaps one of the most compelling advantages of Bayesian nonparametric causal estimation: Aspects of the joint distribution that are identifiable with the observed data, i.e. $f(y_i, a_i, l_i; \omega, \bm v)$, can be modeled flexibly while aspects of the joint distribution that are not identifiable, i.e. $P(\Delta=\delta_i\mid y_i, a_i,l_i;\xi)$, can be modeled parametrically with appropriate priors. The next section will provide a worked example, but notice that this advantage is on display in the bottom-right panel of Figure \ref{fig:tsb_mnar}, which presents the TSB estimators of $E[Y\mid A=a, L=l; \omega, \bm v]$ for each treatment group $a\in\{0,1\}$ under a setting with MNAR - i.e. in which $P(\Delta=1\mid y, a,l;\xi)\neq P(\Delta=1\mid a,l;\xi)$ . Notice that in regions of the $Y-L$ space where there is observed data, the TSB model adapts to the data to fit a flexible curve. However, in regions of the $Y-L$ space where there is substantial missing outcome information, the curve is heavily influenced by our model and priors for the missingness mechanism.

\subsection{A Worked Example with Truncated Stick-Breaking in \stan } \label{sc:tsb_stan_example}

In order to solidify concepts we walk through a worked implementation example in \stan. In this example we have a continuous, real-valued outcome, $Y$, with a single, continuous real-valued confounder, $L$. We specify a TSB mixture model for the joint distribution with $K=10$ components and the following component models:
\begin{equation} \label{eq:mixture_example}
    \begin{split}
     f(y,a,l;\omega, \bm v) & = \sum_{k=1}^K \nu_k f_{Y|AL}(y \mid a, l; \eta_k)f_{A|L}(a \mid l; \gamma_k)f_L(l; \theta_k) \\
     & = \sum_{k=1}^K \nu_k N\Big(y; \eta_{0k} + \eta_{1k}l+\eta_{2k}a, \sigma^2_k\Big)Ber\Big( a; \expit(\gamma_{0k}+\gamma_{1k}l) \Big) N\Big(l;\theta_{0k}, \phi_k\Big)
    \end{split}
\end{equation}
where $\eta_k = (\eta_{0k}, \eta_{1k}, \eta_{2k}, \sigma_k^2)$, $\gamma_k=(\gamma_{0k}, \gamma_{1k})$, and $\theta_k =(\theta_{0k},\phi_k)$. We specify independent priors over these parameters - null-centered Gaussian priors for the regression coefficient and mean parameters and half-normal priors on the variance parameters. With independent priors, $g_\Omega(\omega)$ is the the product of these densities. 

\begin{figure}[h!]
    \centering
    \includegraphics[scale=.45]{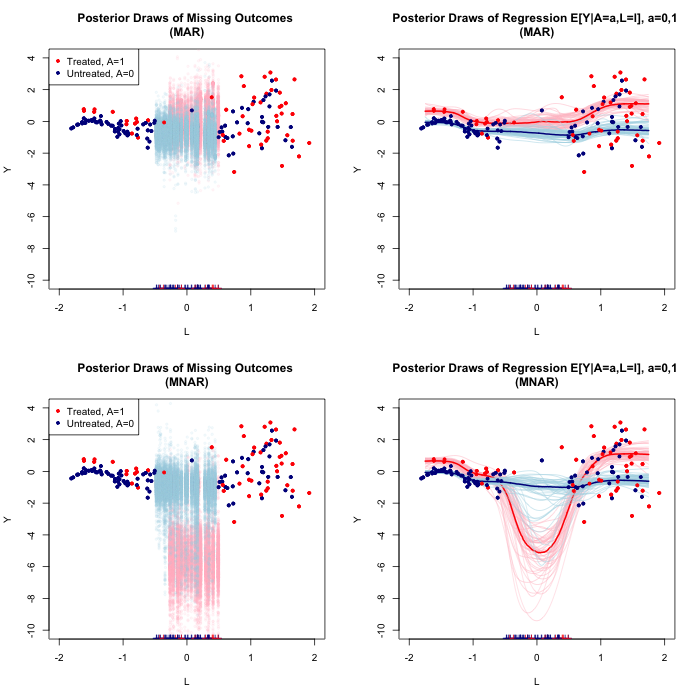}
\caption{ Posterior Inference with TSB model and missing outcomes. Red/blue points depict treated/untreated subjects. Treated/Untreated subjects with missing outcomes are represented as a tick mark on the x-axis at the location of their observed $L$ value. Top Row: Bayesian inference under the model described in Section \ref{sc:tsb_stan_example} with $\xi_3=0$ - i.e. a strong prior belief that outcome missingness is at-random (MAR). The left panel shows, at each tick, several posterior draws of the missing outcome for that subject. Draws for treated/untreated subjects are shown in light red/light blue. The spread reflects posterior uncertainty about that subject's outcome value. The right panel shows the posterior mean regression line for treated/untreated subjects in dark red/blue. Several posterior draws of the regression line are overlaid in lighter shades to visualize posterior uncertainty. Bottom Row: Same analysis as in the top row but with a prior belief in MNAR with $\xi_3=-1$. This is a prior belief that treated patients with missing outcomes had lower than average outcomes. Notice that the posterior draws of the missing outcome values for treated subjects are now shifted down in the right panel. Consequently, the regression line for treated subjects now has a dip to reflect this prior belief in MNAR outcomes.}
\label{fig:tsb_mnar}
\end{figure}

As in Section \ref{sc:ex_mnar_parametric}, we will model the missingness mechanism as a logistic regression with sensitivity parameters $\xi=(\xi_0, \xi_1,\xi_3)$

$$  P(\Delta =1 \mid y, a, l;\xi) = \expit\big ( \xi_0 + \xi_1a + \xi_3 (a\cdot y) \big )$$

Substituting the particular forms of $P(\Delta =1 \mid y, a, l;\xi)$ and \eqref{eq:mixture_example} into Equation \eqref{eq:mixture_post_missing} and taking the log yields the log unnormalized joint posterior for the model that can be passed to \stan in the \cd{model} block is

\begin{equation}
    \begin{split}
        \log \tilde f(D^M, \xi, \omega, \bm v \mid D^o) & \propto \log g_\Omega(\omega) + \log f_{\bm V} (\bm v) \\ 
        & \ \ \ \  + \sum_{i=1}^n \log  Ber\Big(\delta_i;\expit( \xi_0 + \xi_1a + \xi_3 a y )\Big)\\
        & \ \ \ \  + \sum_{i=1}^n \log\Big( \sum_{k=1}^K \nu_k N\Big(y_i; \eta_{0k} + \eta_{1k}l_i+\eta_{2k}a_i, \sigma^2_k\Big) Ber\Big( a_i; \expit(\gamma_{0k}+\gamma_{1k}l_i) \Big)N(l_i;\theta_{0k}, \phi_k)\Big)
    \end{split}
\end{equation}

Since implementation is similar to previous examples, we do not provide an excerpt here and refer interested readers to the full code on the companion GitHub repository. Figure \ref{fig:tsb_mnar} displays the posterior inference under two priors beliefs: a belief that $\xi_3 \sim \bar \delta_0$ (top row) and a belief that $\xi_3 \sim \bar \delta_{-1}$ (bottom row). The former encodes a strong prior belief in MAR, while the latter encodes a strong prior belief in MNAR. The top-left panel of Figure \ref{fig:tsb_mnar} shows the posterior draws of $D^M=\{y_i\}_{i=n_o+1}^{n}$ and we see that the posterior draws of these missing outcome are similar across treated and untreated subjects. This is an artifact of our prior belief in MAR. The top-right panel of the figure shows that posterior draws of the regression line obtained by plugging in the $\omega^{(m)},\bm v^{(m)}$ into Equation \eqref{eq:tsb_outcome}, $E[Y\mid A=a, L=l;\omega^{(m)},\bm v^{(m)}]$ for each $a\in\{0,1\}$. We see that the TSB model is able to capture the complex dependence between $Y$ and $L$ within each treatment group.

Now for the MNAR setting in the bottom row of Figure \ref{fig:stan_mnar}, note that $\xi_3 \sim \bar \delta_{-1}$ encodes a belief that treated patients with a one-standard deviation above average outcome have a lower odds  (odds ratio $\exp(-1) \approx .38)$ of missing the outcome relative to those with average outcome level. Equivalently, the missing outcome values among treated subjects are likely lower than the observed outcome values among the treated. Accordingly, we see in the bottom-left panel of the figure that the posterior draws of the missing outcome differ across treated and untreated subjects. The posterior draws for the treated subjects are much lower and this is an artifact of our MNAR prior $\xi_3 \sim \bar \delta_{-1}$. Consequently, the posterior regression line for the treated group in the bottom-right panel now dips to account for this prior belief in MNAR. 

This ability to 1) be data-adaptive/flexible in regions of the data space with many observations and 2) incorporating prior beliefs in regions of the data space without observations 3) while quantifying uncertainty that reflects both 1) and 2) is an attractive and distinguishing feature of Bayesian sensitivity analysis in causal inference.

\section{Discussion}

Sensitivity analysis is a large area of research and, consequently, there are many ways to assess sensitivity. For example, some methods for assessing sensitivity to unmeasured confounding using ``confounding functions'' or ``bias functions'' \cite{Hu2022, Oganisian2021a} while others use a latent unmeasured confounder approach as in Example 2 \cite{Hu2022}. Popiular frequentist approaches sometimes correspond to special cases of the fully Bayesian approach discussed here. For example, in the context of Example 2, an E-value approach \cite{VanderWeele2017} involves setting $\xi_1=\xi_2=\xi^*$ and finding the minimum such $\xi^*$ that would ``explain away'' an estimated effect. We presented a ``missing data'' framing as it provides a unified way of conceptualizing violations to a several assumptions from from exposure misclassification (Example 1) and unmeasured confounding (Example 2) to missing outcome information (Examples 3 and 4). 

Throughout, we mostly focus on sensitivity analyses using point-mass (i.e. degenerate) priors on sensitivity parameters and traced out how posteriors change across a grid of possible points. For instance, in Example 4, we set the prior to a point-mass at $\xi_3=1$, $\xi_3\sim \bar\delta_{-1}$. However, the Bayesian approach allows for non-degenerate priors such as $\xi_3\sim N(0,1)$. This encodes a prior belief that is centered on MAR, but with symmetric uncertainty about direction/magnitude of a violation. This approach yields a single posterior for $\Psi$ that averages over prior uncertainty about $\xi_3$, rather than tracing out a series of posteriors across a grid.

A key contribution of this paper was to present code and math side-by-side to help readers build implementation intuition for their own sensitivity analyses. Perhaps contrary to popular thought, we hope that the paper conveys the fact that practical and interesting Bayesian sensitivity analyses can be done with publicly available software. Although many \stan wrappers exist, we focus on a ``from scratch'' approach since it explicitly connects the mathematics/structure of the posterior with what is coded in the \cd{model} block in \stan. Similarly, we present code that is a ``literal'' translation of the math rather than the most efficient \stan programming practice. This is, again, a didactic choice aimed at helping the reader develop intuition and knowledge that generalizes beyond the examples covered in this paper.

Lastly, we hope this paper makes a convincing case that substantive, application-specific expertise is required to both constructively critique the plausibility of an assumption as well as to respond to such critiques. For example, simply claiming ``there may be an unmeasured confounder'' is a lazy critique as this is trivially true in any application. A constructive criticism would posit a specific clinical feature that has gone unmeasured as well as positing a direction/magnitude of a resulting bias. Such a criticism can then be constructively addressed using principled sensitivity analysis methods like the ones illustrated here.

%\backmatter

\bmsection*{Acknowledgments}
This work was partially funded by Patient-Centered Outcomes Research Institute (PCORI) contracts ME-2023C1-31348, ME-2021C3-2494, and National Institutes of Health (NIH) grant 5R01AI167694-02.

\bmsection*{Financial disclosure}

None reported.

\bmsection*{Conflict of interest}

The authors declare no potential conflict of interests.

\bibliography{wileyNJD-AMA}

\bmsection*{Supporting information}

Additional supporting information may be found in the online version of the article at the publisher’s website.

%\appendix
%\bmsection{Example Appendix\label{app1}}
%\vspace*{12pt}

\end{document}